\documentclass[journal]{IEEEtran}

\usepackage{cite}
\usepackage{graphicx}
\usepackage{amsmath,amssymb,bm}
\usepackage{booktabs}
\usepackage{multirow}
\usepackage[caption=false,font=footnotesize]{subfig}
\usepackage{array}
\usepackage{tabularx}
\usepackage{makecell}
\usepackage{xcolor}
\usepackage[colorlinks=true,linkcolor=blue,citecolor=blue,urlcolor=blue]{hyperref}

\begin{document}

\title{JointHRRP-Net: A Statistically Constrained Decoupling Network for Joint Target and Jamming Recognition in Composite Jamming}

\author{Yunfei~Zhao,
        Mei~Liu,
        Shuowei~Liu,
        Xunzhang~Gao,  
        and~Yujie~Zhou%
\thanks{This work was supported in part by the National Natural Science Foundation of China under Grant 61921001, and the Hunan Provincial Graduate Student Research Innovation Program under Grant CX20250020. (Corresponding author: Xunzhang Gao.)}%
\thanks{Yunfei Zhao, Mei Liu, Shuowei Liu, Xunzhang Gao, and Yujie Zhou are with the College of Electronic Science and Technology, National University of Defense Technology, Changsha $410073$, China. (e-mail: zhao\_yunfei@nudt.edu.cn; liumei18@nudt.edu.cn; liushuowei@nudt.edu.cn; gaoxunzhang@nudt.edu.cn; zhouyujie\_02@nudt.edu.cn)}}

\markboth{Preprint}{Zhao \MakeLowercase{\textit{et al.}}: JointHRRP-Net for Joint Target and Jamming Recognition}

\maketitle

{\footnotesize
\textit{This manuscript has been submitted to the IEEE for possible publication.}
\par\vspace{0.5em}}

\begin{abstract}
High-resolution range profile (HRRP)-based radar automatic target recognition suffers from severe performance degradation in composite jamming environments. Active jamming introduces suppression- and deception-related components into the received range profile. After pulse compression, these components are coupled with target echoes in the HRRP domain, making target-related scattering peaks difficult to distinguish and weakening feature separability. To address this problem, this paper proposes JointHRRP-Net, a unified framework for joint target--jamming recognition. A statistically constrained decoupling module is first developed to generate target-dominant and jamming-dominant latent branches from the mixed HRRP representation. Correlation-guided statistical constraints are imposed to suppress redundant cross-branch information and alleviate target--jamming feature entanglement. A multi-scale temporal encoding module is then designed to model local scattering structures and long-range range-cell dependencies, followed by a dual-expert decision module for single-label target classification and multi-label jamming classification. Experiments under diverse signal-to-jamming ratio (SJR) and signal-to-noise ratio (SNR) levels demonstrate that JointHRRP-Net outperforms representative baseline methods in both target recognition and composite jamming recognition. Open-set evaluation further shows that the learned target representation remains discriminative for unknown-target rejection. These results demonstrate the effectiveness and robustness of JointHRRP-Net in composite jamming scenarios.
\end{abstract}

\begin{IEEEkeywords}
Radar automatic target recognition, active jamming, high-resolution range profile, representation decoupling, joint target--jamming recognition, open-set recognition.
\end{IEEEkeywords}

\section{Introduction}

\IEEEPARstart{R}{ADAR} automatic target recognition (RATR) is an important technology for battlefield situational awareness and intelligent decision-making~\cite{liu2025TAESDeepGHMMHRRP,liu2024PKGuidedHRRP,zhou2025TAESDualPolHRRP,tao2025TAESOpenSetAAV}. Among various radar representations, the high-resolution range profile (HRRP) provides a compact one-dimensional description of the target scattering distribution along the radar line of sight. Owing to its convenient acquisition and efficient processing, HRRP-based recognition has important engineering value for rapid radar target recognition. Compared with synthetic aperture radar (SAR) image-based recognition, HRRP-based target recognition avoids long coherent integration and usually has lower data acquisition latency, making it suitable for agile radar platforms and real-time deployment~\cite{zhou2025TAESDualPolHRRP,tao2025TAESOpenSetAAV}.

\begin{figure}[!htbp]
\centering
\includegraphics[width=0.3\textwidth]{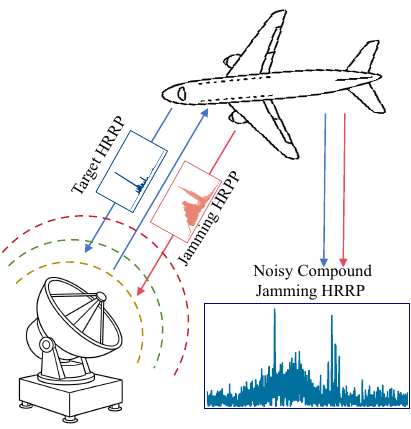}
\caption{Formation mechanism of an HRRP under composite jamming scenarios. After pulse compression, target scattering components, active jamming components, and noise components are superposed in the received range-profile domain. The coupled composite return makes target-related scattering peaks difficult to distinguish, thereby degrading subsequent radar target recognition.}
\label{fig:hrrp_formation}
\end{figure}

However, with the rapid development of digital radio frequency memory (DRFM) devices, active jamming techniques have become increasingly sophisticated, posing serious challenges to radar perception in complex electromagnetic environments~\cite{wang2025JSTARSRadarJammingSurvey,zhao2026JRPriorGuidedCompositeInterference}. In practical scenarios, suppression jamming, deception jamming, target echoes, and thermal noise may coexist within the same radar observation~\cite{zheng2025TAESD2AFDeceptionJamming,wuship}. After pulse compression, these components are superposed in the range-profile domain. As a result, target scattering peaks may become difficult to distinguish because they are mixed with jamming components in the composite return. Fig.~\ref{fig:hrrp_formation} illustrates the formation mechanism of an HRRP sample under composite jamming. Such target--jamming coupling impedes target information extraction and consequently degrades HRRP-based target recognition performance.

Deep learning has greatly promoted HRRP target recognition, open-set recognition, and robust feature learning \cite{pan2022TGRSHRRPStacked,CHEN2022,liu2024PKGuidedHRRP,xia2022hrrpopenset,xiaopen,du2008TSPHRRPStatistical,guo2020TSPVariationalHRRP,liao2018AccessConcatHRRP,wang2023TGRSCorruptedHRRP,gaotrans,msdpnet,zhou2025TAESDualPolHRRP}. CNN-based feature extraction, recurrent temporal modeling, and attention mechanisms have been widely used to capture local range-cell patterns and long-range sequential dependencies \cite{pan2022TGRSHRRPStacked,CHEN2022,vaswani2017attention}. Scattering-prior-guided learning, joint denoising and recognition, graph-structured modeling, and domain adaptation have further improved HRRP recognition robustness under noisy, data-limited, and domain-shifted conditions \cite{li2023TGRSHRRPCVAE,liu2024PKGuidedHRRP,chen2024HRRPGraphNet,chen2025HRRPGraphNetPP,liu2022IntegratedDenoiseHRRP,zhou2024LGRSDomainAdaptiveHRRP,su2023LGRSScatteringCenterSeqHRRP}. Nevertheless, most existing HRRP target recognition methods still assume target-only or mildly corrupted inputs. Open-set HRRP protocols also tend to evaluate unknown-target rejection under relatively clean or target-dominated conditions. Their effectiveness remains insufficiently explored when target scattering structures are coupled with multiple active jamming components in the same pulse-compressed profile.

To support anti-jamming radar perception, active jamming recognition and suppression have also been widely investigated. For single jamming categories, weighted ensemble CNNs and transfer learning strategies have been introduced to enhance deception-jamming discrimination, while time--frequency self-attention with knowledge distillation has been used to reduce the dependence on large-scale labeled jamming samples \cite{lv2022TGRSDeceptionJamming,luo2023TGRSFewShotJamming}. As practical electronic attack becomes more complex, several works have extended jamming recognition from single-label classification to composite and multi-label classification. Deep multi-label learning, JRNet, and lightweight multi-label networks have been developed to recognize multiple jamming components in composite jamming scenarios \cite{Zhu2020AMR,qujrnet,2024CompoundJammingRecognition}. Although these studies show promising performance for jamming-oriented analysis, most of them focus on jamming components only and do not explicitly consider HRRP observations that simultaneously contain target echoes, composite jamming, and noise.

Overall, HRRP recognition under composite jamming remains challenging in the following aspects.

\begin{enumerate}
\item Feature entanglement: Target-dominant and jamming-dominant components are strongly coupled after matched filtering. Features extracted from mixed HRRP profiles may therefore become unstable or biased toward jamming characteristics.

\item Output heterogeneity: HRRP target recognition is typically formulated as single-label classification, whereas composite jamming recognition requires multi-label prediction because multiple jamming mechanisms may coexist in one observation. A homogeneous classifier is therefore insufficient for joint target--jamming decision-making.

\item Robust open-set recognition: Low-SNR, low-SJR, and unknown target categories further increase recognition difficulty. The model is required to maintain accurate closed-set discrimination while preserving separable representations for reliable rejection.
\end{enumerate}

These challenges indicate the necessity of a unified framework that can jointly model target-related and jamming-related information while maintaining effective feature separation under complex electromagnetic interference.

To address these issues, this paper proposes JointHRRP-Net, a unified HRRP target and jamming recognition network under composite jamming. Instead of directly extracting features from the mixed HRRP profile, JointHRRP-Net introduces a statistically constrained decoupling strategy to separate target-dominant and jamming-dominant components in latent space. The decoupled representations are then processed by a multi-scale temporal encoder and dual-expert decision heads. This design avoids a purely cascaded suppression-recognition pipeline and provides an integrated solution for target recognition, composite jamming recognition, and open-set target analysis in contested electromagnetic environments.

The main contributions of this paper are summarized as follows.

\begin{enumerate}
\item A statistically constrained decoupling module is proposed to generate target-dominant and jamming-dominant latent branches from mixed HRRP profiles. Correlation-guided constraints are imposed between the two branches to suppress redundant cross-branch information and alleviate feature entanglement caused by target--jamming coupling.

\item A multi-scale temporal encoding module is designed to enhance HRRP representation under low-SNR and low-SJR conditions. The module captures local scattering patterns and long-range range-cell dependencies within the decoupled branches, improving the robustness of target and jamming feature learning.

\item A dual-expert decision module is developed to jointly support single-label target recognition and multi-label jamming recognition. The task-specific heads accommodate heterogeneous output structures within one framework, which improves joint recognition stability under composite jamming and open-set conditions.
\end{enumerate}

\section{Related Work}

HRRP target recognition and radar active jamming recognition have been extensively investigated in recent years. Although both topics are closely related to radar perception in complex electromagnetic environments, they have mainly been developed along relatively independent technical routes. This section reviews related studies on HRRP target recognition, radar active jamming recognition, and joint target--jamming recognition under complex electromagnetic interference.

\subsection{HRRP Target Recognition}

Early HRRP target recognition methods mainly relied on hand-crafted descriptors, statistical modeling, template matching, or sparse representation. Tree-structured hierarchical feature selection provides a classical solution for ground target recognition \cite{liutree}. Statistical-template modeling, scatterer-pattern analysis, variational inference, and feature concatenation further enrich the traditional HRRP recognition framework \cite{du2008TSPHRRPStatistical,feng2016ICASSPHRRPScatterer,guo2020TSPVariationalHRRP,liao2018AccessConcatHRRP}. These methods are closely related to physical scattering characteristics and therefore provide a certain degree of interpretability. However, their performance is usually sensitive to aspect variation, noise disturbance, and distribution shift, which limits their reliability under strongly corrupted HRRP observations.

\begin{figure}[!htbp]
  \centering
  \subfloat[MSDP-Net without jamming]
  {\includegraphics[width=0.5\columnwidth]{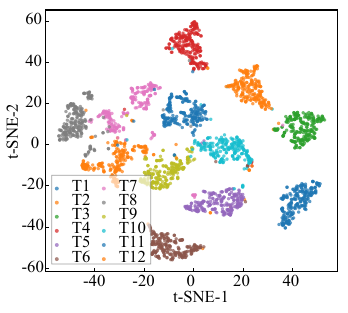}
  \label{fig:tsne_target_only_no_noise_msdp_net}}%
  \subfloat[MSDP-Net under strong jamming]
  {\includegraphics[width=0.5\columnwidth]{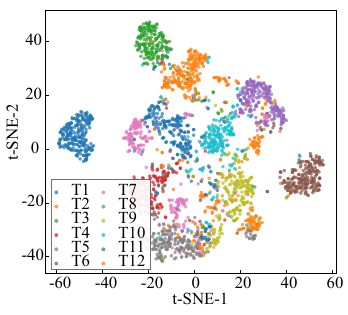}
  \label{fig:tsne_with_interf_no_noise_msdp_net}}\\

  \subfloat[DPFFN without jamming]
  {\includegraphics[width=0.5\columnwidth]{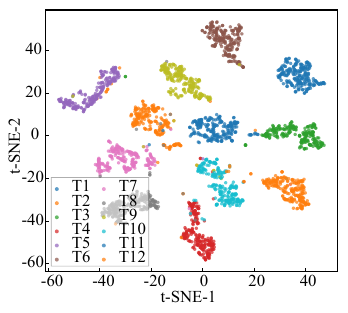}
  \label{fig:tsne_target_only_no_noise_dpffn}}%
  \subfloat[DPFFN under strong jamming]
  {\includegraphics[width=0.5\columnwidth]{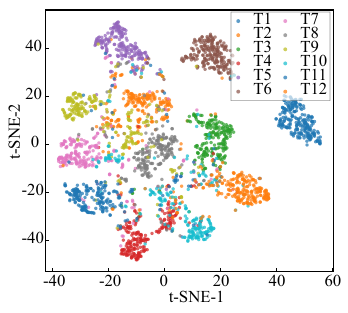}
  \label{fig:tsne_with_interf_no_noise_dpffn}}

  \caption{t-SNE visualization under noise-free conditions. The left column shows feature distributions without jamming, whereas the right column shows feature distributions under strong composite jamming.}
  \label{fig:tsne_interference_comparison}
\end{figure}

With the development of deep learning, HRRP recognition has gradually shifted from manual feature engineering to end-to-end representation learning. CNN-based, recurrent, and attention-based networks have been used to model local range-cell structures and sequential dependencies \cite{pan2022TGRSHRRPStacked,CHEN2022}. Residual--inception structures, multistatic recognition networks, corrupted-HRRP modeling, and transformer-based representation learning further improve the expressive capability of deep HRRP features \cite{guo2019AccessResidualInceptionHRRP,wang2016RadarConfMultistaticHRRP,wang2023TGRSCorruptedHRRP,gaotrans}. More recently, MSDP-Net introduces multi-domain HRRP feature learning through multi-scale spatial convolutions, adaptive spectral gating, and hierarchical semantic fusion \cite{msdpnet}, while DPFFN exploits dual-polarization feature fusion for radar recognition \cite{zhou2025TAESDualPolHRRP}. These methods have achieved promising closed-set recognition performance. Nevertheless, most of them are still designed for target-only or mildly disturbed HRRP inputs, and they do not explicitly consider the pulse-compressed mixture of target echoes and active jamming. As shown in Fig.~\ref{fig:tsne_interference_comparison}, strong composite jamming makes the feature distributions more scattered and overlapped, indicating that active jamming severely weakens the separability of target HRRP representations.

Several studies have further improved HRRP recognition under data-limited, noisy, and domain-shifted conditions. Scattering-prior-guided supervised contrastive learning reduces intra-class dispersion and enhances inter-class separation \cite{liu2024PKGuidedHRRP}. HRRPGraphNet and HRRPGraphNet++ encode range-cell relations through graph structures and dynamic graph attention, thereby improving few-shot and low-SNR recognition \cite{chen2024HRRPGraphNet,chen2025HRRPGraphNetPP}. Continual learning, joint denoising and recognition, sequential scattering-center modeling, gated recurrent fusion, domain adaptation, and cognitive waveform-assisted recognition have also been investigated to improve HRRP robustness from different perspectives \cite{li2023TGRSHRRPCVAE,liu2022IntegratedDenoiseHRRP,su2023LGRSScatteringCenterSeqHRRP,zeng2022LGRSMICGRUHRRP,zhou2024LGRSDomainAdaptiveHRRP,warnke2023TAESCognitiveHRRP,liu2025TAESDeepGHMMHRRP,li2024TAESFewShotShrinkageHRRP}. However, their experimental settings are still mainly based on target-dominated signals or front-end denoised profiles. Strong in-pulse composite jamming in the HRRP domain has not been sufficiently investigated.

Open-set HRRP recognition is another important topic because practical radar systems may encounter unknown targets. Surrounding prototype loss and extreme-value modeling enlarge the separation between known and unknown classes \cite{xia2022hrrpopenset,xiaopen}. Class-imbalanced open-set recognition, motion-sequence modeling, threshold-insensitive scoring, consistency-aware prototypes, systematic open-set frameworks, and multivariate extreme-value boundaries have further extended HRRP recognition from closed-set classification to open-set recognition \cite{ZHANGopen,zhang2025TAESHRRPSeqNet,tao2025TAESOpenSetAAV,chen2025TAESECAPL,li2025TAESOSFSM,li2025TAESMVEVOpenSetHRRP}. However, many of these methods rely on complex open-set mechanisms and are usually evaluated under relatively clean or simplified conditions. Recent studies also indicate that a strong closed-set classifier can provide a competitive basis for open-set adaptation when the learned feature space is sufficiently discriminative \cite{vaze2022openset}. Therefore, robust closed-set feature learning remains a fundamental requirement for open-set target analysis under composite jamming.

\subsection{Radar Active Jamming Recognition}

Radar active jamming recognition aims to recognize the jamming type or jamming composition from radar observations. Existing methods can be broadly divided into single-category jamming recognition, composite or multi-label jamming recognition, and jamming-oriented suppression or countermeasure analysis. For single-category recognition, weighted ensemble CNNs and transfer learning strategies have been used to improve deception-jamming discrimination under limited training conditions \cite{lv2022TGRSDeceptionJamming}. Time--frequency self-attention combined with global knowledge distillation further enhances few-shot jamming recognition by transferring useful prior knowledge from larger models \cite{luo2023TGRSFewShotJamming}. Feature-distance metric learning has also been explored for imbalanced SAR jamming recognition \cite{cen2024TGRSSARJammingRecognition}. These methods improve the feature separability of individual jamming modes, but they are not designed for observations containing multiple simultaneous jamming mechanisms.

To recognize composite jamming, several studies have extended single-label classification to multi-label or multi-component prediction. Deep multi-label learning provides a direct solution for composite jamming classification \cite{Zhu2020AMR}. JRNet focuses on compound suppression jamming recognition, while complex-valued CNNs, prototype-based class representations, attention-based class decoupling, and lightweight multi-feature fusion networks further improve time-domain or resource-constrained composite jamming recognition \cite{qujrnet,meng2023MLCRCVCNN,zha2026CompoundJammingMLFF,2024CompoundJammingRecognition}. These methods explicitly model the non-exclusive relationship among different jamming types and have shown promising performance in jamming-oriented scenarios.

Jamming suppression and countermeasure studies provide another route for improving radar perception. Attention-embedded UNet-A enables range-cell-level recognition and suppression of interrupted-sampling repeater jamming (ISRJ), but such methods usually require dense range-cell annotations and their robustness under composite jamming remains difficult to guarantee \cite{wuUNet}. Fractional-domain local--global feature fusion with attention strengthens the representation of salient jamming characteristics, and ISRJ counter-waveform design or mismatched filtering can reduce the influence of specific jamming mechanisms \cite{zhoujam2023,zhou2022TGRSISRJWaveform,zhou2023TGRSISRJCountermeasure}. These studies are effective for jamming analysis and countermeasure design. However, most of them either predict jamming types only or treat jamming suppression as a separate preprocessing step before target classification. The target identity and jamming composition are rarely modeled as coupled outputs within a unified HRRP recognition framework.

\subsection{Research Gap in Joint Target--Jamming Recognition}

The above studies show that HRRP target recognition and radar active jamming recognition have both achieved considerable progress. Nevertheless, they are still mainly modeled as two isolated tasks. This assumption is limited in practical composite jamming scenarios, where the measured HRRP profile may contain target scattering, active jamming, and noise simultaneously. A cascaded suppression-recognition pipeline is vulnerable to front-end suppression errors, which may remove useful target scatterers or leave residual false components. Directly applying a simple multitask classifier to the raw mixed profile is also insufficient, because the shared representation can be dominated by the stronger component, especially under low-SNR or low-SJR conditions.

Feature-level disentanglement provides a possible route for addressing this problem. Denoising-guided disentanglement has been discussed in general radar signal classification, where useful components and nuisance factors are separated before decision-making \cite{zhang2022DisentangledIntrapulse}. However, pulse-compressed aircraft HRRP recognition under composite jamming is more demanding. It requires target-oriented recognition, multi-label jamming recognition, and potential unknown-target rejection within the same observation. Meanwhile, the model should preserve target scattering structures while describing jamming composition.

Therefore, a clear gap remains in unified HRRP recognition under composite jamming. An effective model should satisfy three requirements. First, target-dominant and jamming-dominant factors should be decoupled at the representation level, instead of relying only on raw-profile classification. Second, task-specific decision heads are needed because target recognition and composite jamming recognition follow different output structures. Third, robust and separable representations should be maintained under low-SNR, low-SJR, and open-set conditions. To address these requirements, this paper develops JointHRRP-Net through an integrated design of statistical decoupling, multi-scale temporal encoding, and dual-expert decision-making.

\section{Signal Model and Dataset Construction}

This section formalizes the generation of pulse-compressed HRRPs in the presence of target echoes and multiple active jamming components, thereby providing a physical basis for constructing mixed HRRP signals and for the target--jamming representation decoupling in the proposed network. 

\subsection{Formation of HRRP Under Composite Jamming}

Let $r(t)$ denote the received radar signal. In the presence of target echo, suppression jamming, deception jamming, and thermal noise, the received signal can be expressed as
\begin{equation}
\begin{aligned}
r(t)= {} & s_{\mathrm{tar}}(t)
+\sum_{i=1}^{I}J_{\mathrm{sup},i}(t)
+\sum_{k=1}^{K}J_{\mathrm{dec},k}(t)
+n(t)
\end{aligned}
\label{eq:received}
\end{equation}
where $s_{\mathrm{tar}}(t)$ denotes the true target echo, $J_{\mathrm{sup},i}(t)$ denotes the $i$th suppression jamming component, $J_{\mathrm{dec},k}(t)$ denotes the $k$th deception jamming component, $I$ and $K$ are the numbers of suppression and deception jamming components, respectively, and $n(t)$ is additive Gaussian noise.

After pulse compression with the matched filter $h(t)$, the received signal is transformed into the HRRP domain as
\begin{equation}
\begin{aligned}
y(t)= {} & r(t)\ast h(t) \\
= {} & s_{\mathrm{hrrp}}(t)
+\sum_{i=1}^{I}J'_{\mathrm{sup},i}(t)
+\sum_{k=1}^{K}J'_{\mathrm{dec},k}(t)
+n'(t)
\end{aligned}
\label{eq:pulsecompressed}
\end{equation}
where $\ast$ denotes convolution, $s_{\mathrm{hrrp}}(t)$ is the clean target HRRP, $J'_{\mathrm{sup},i}(t)$ and $J'_{\mathrm{dec},k}(t)$ are the compressed suppression and deception jamming components, respectively, and $n'(t)$ is the compressed noise term.

Consequently, the target scattering response, active jamming components, and noise are superposed in the HRRP domain after pulse compression. This superposition leads to a more complex HRRP feature distribution than that in the clean target-only case.

\subsection{Representative Jamming Models}

Composite jamming usually contains both suppression-oriented and deception-oriented components. Among different active jamming modes, coherent DRFM-based waveforms are particularly effective because they can generate false structures or broaden the jamming distribution in the HRRP domain. In this paper, four representative jamming modes are considered, including chopping and interleaving (C\&I) jamming, interrupted-sampling repeater jamming (ISRJ), spread modulated spectrum perturbation (SMSP), and noise convolution jamming (NCJ). For each training or test sample, a non-empty subset of these four jamming types is uniformly selected. Therefore, one to four active jamming types may coexist in one observation, leading to $2^4-1=15$ possible multi-label jamming combinations.

Let $x(t)$ denote the intercepted radar waveform used for jamming generation. The analytical forms of the four jamming modes are introduced as follows.

For C\&I jamming, the intercepted waveform is first chopped into sub-pulses and then rearranged across different temporal slots. The corresponding signal is expressed as
\begin{equation}
\begin{aligned}
s_{\mathrm{C\&I}}(t)= {} & A\sum_{r=0}^{N_{\mathrm{R}}-1}\sum_{c=0}^{N_{\mathrm{C}}-1}
x\!\left(t-\frac{rT}{N_{\mathrm{C}}N_{\mathrm{R}}}\right) \\
& \cdot \left[
u\!\left(t-\frac{cT}{N_{\mathrm{C}}}\right)
-u\!\left(t-\frac{cT}{N_{\mathrm{C}}}-\frac{T}{N_{\mathrm{R}}}\right)
\right]
\end{aligned}
\label{eq:ci}
\end{equation}
where $u(\cdot)$ denotes the unit step function, $N_{\mathrm{C}}$ is the number of chopped sub-pulses, and $N_{\mathrm{R}}$ is the number of interleaving slots.

For ISRJ, the jammer intermittently samples the intercepted radar waveform and retransmits the sampled slices with multiple delays \cite{zhou2022TGRSISRJWaveform,zhou2020LSPISRJWaveform,liu2018TAPHRRPInterrupted}. Its signal can be written as
\begin{equation}
\begin{aligned}
s_{\mathrm{ISRJ}}(t)= {} & A\sum_{p=1}^{P}\sum_{q=1}^{Q}
\mathrm{rect}\!\left(\frac{t-\alpha(p,q)T_{\mathrm{I}}}{T_{\mathrm{I}}}\right)
x(t-qT_{\mathrm{I}})
\end{aligned}
\label{eq:isrj}
\end{equation}
where $P$ is the number of sampling operations, $Q$ is the number of forwarding operations, $\alpha(p,q)=(p-1)(Q+1)+q$ is the delay coefficient, $T_{\mathrm{I}}$ is the sampling width, $A$ is the jamming gain, and $\mathrm{rect}(\cdot)$ denotes the rectangular window.

For SMSP, the jamming waveform is generated by assigning different carrier frequencies and chirp slopes to different temporal slices. It is formulated as
\begin{equation}
\begin{aligned}
s_{\mathrm{SMSP}}(t)= {} & A\sum_{m=1}^{M_{\mathrm{S}}}
\exp\!\left(j2\pi\left[f_m t+\frac{k_m t^2}{2}\right]\right), \\
& t\in\left[\frac{(m-1)T}{M_{\mathrm{S}}},\frac{mT}{M_{\mathrm{S}}}\right]
\end{aligned}
\label{eq:smsp}
\end{equation}
where $M_{\mathrm{S}}$ is the number of slices, $T$ is the radar pulse width, $T/M_{\mathrm{S}}$ is the duration of each slice, $f_m$ is the carrier frequency of the $m$th slice, and $k_m$ is the chirp slope.

For NCJ, the intercepted waveform is convolved with a noise kernel in the time domain. This operation enlarges the effective compressed energy distribution of the jamming signal. The NCJ signal is given by
\begin{equation}
s_{\mathrm{NCJ}}(t)=A\left[x(t)\ast w(t)\right]
\label{eq:ncj}
\end{equation}
where $w(t)$ is the noise convolution kernel.

\begin{figure}[!htbp]
  \centering
  \subfloat[C\&I]
  {\includegraphics[width=0.5\columnwidth]{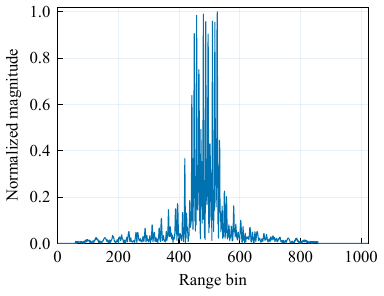}
  \label{fig:jam_hrrp_ci}}%
  \subfloat[ISRJ]
  {\includegraphics[width=0.5\columnwidth]{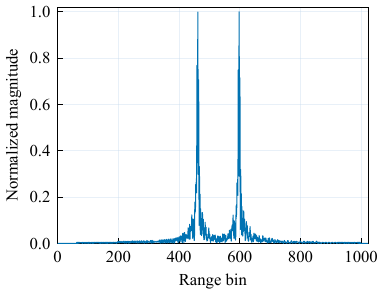}
  \label{fig:jam_hrrp_isrj}}\\

  \subfloat[SMSP]
  {\includegraphics[width=0.5\columnwidth]{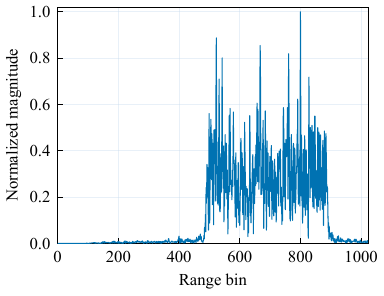}
  \label{fig:jam_hrrp_smsp}}%
  \subfloat[NCJ]
  {\includegraphics[width=0.5\columnwidth]{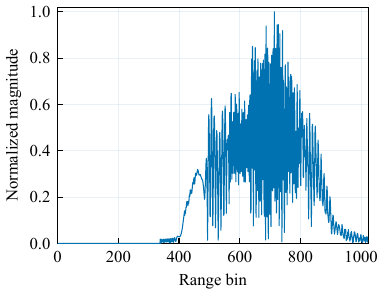}
  \label{fig:jam_hrrp_ncj}}

  \caption{Illustrative range profiles of the four jamming templates adopted in this paper. Each subplot depicts the normalized magnitude versus range bin of a typical pulse-compressed jamming sample.}
  \label{fig:jam_hrrp_four_types}
\end{figure}

From \eqref{eq:ci}--\eqref{eq:ncj} and the illustrative range profiles in Fig.~\ref{fig:jam_hrrp_four_types}, jamming signals are generally generated by modulating the transmitted radar waveform. Consequently, their analytical expressions and waveform structures are relatively deterministic, leading to highly regular statistical distributions. By contrast, the HRRP is mainly governed by the spatial distribution of target scattering centers. Its amplitude and phase responses vary dynamically with the aspect angle and target geometry, resulting in relatively weak statistical regularity in the target range profile. This discrepancy motivates the separation of jamming-dominated regular patterns from target-dominated scattering structures.

\subsection{Dataset Construction}
\label{subsec:dataset_construction}

The target data used in this paper are constructed from two benchmark HRRP datasets: an X-band aircraft electromagnetic simulation dataset and a widely used measured dataset. The simulation dataset contains twelve aircraft categories, denoted as A1--A12, and the measured dataset contains three aircraft categories, denoted as M1--M3. Owing to their rich scattering characteristics, polarization information, and high-fidelity radar measurements, these datasets have been widely used for evaluating HRRP target recognition algorithms \cite{liu2024PKGuidedHRRP,liu2025TAESDeepGHMMHRRP,LIU2025109876}.

To facilitate subsequent joint recognition of target and jamming information, all HRRP samples are zero-padded along the range dimension to a uniform length of 1024. This preprocessing step not only ensures a consistent input size across the two datasets, but also preserves more complete range-profile information for both target scattering structures and active jamming components. In particular, since the jamming-induced distortions and the target scattering patterns may occupy different range cells, the longer range window helps retain more discriminative details for composite sample construction and downstream recognition.

Based on the four jamming models in \eqref{eq:ci}--\eqref{eq:ncj}, the corresponding jamming components are generated under the same radar system parameters as those used for the clean HRRP templates in the two datasets. Then, pulse compression is performed to construct composite jamming-contaminated HRRP samples. This procedure enables a realistic simulation of range-profile distortion caused by active jamming in practical electronic countermeasure scenarios.

Let $\mathbf{t}$, $\mathbf{i}$, and $\mathbf{n}$ denote the clean target HRRP template, composite jamming profile, and additive Gaussian noise vector within the effective HRRP range window, respectively. Their energies are defined as
\begin{equation}
E_{\mathrm{tar}}=\lVert\mathbf{t}\rVert_2^2,\quad
E_{\mathrm{jam}}=\lVert\mathbf{i}\rVert_2^2,\quad
E_{\mathrm{n}}=\lVert\mathbf{n}\rVert_2^2.
\end{equation}
The signal-to-jamming ratio (SJR) and signal-to-noise ratio (SNR) are calculated in decibels as
\begin{equation}
\begin{aligned}
\mathrm{SJR} &= 10\log_{10}(E_{\mathrm{tar}}/E_{\mathrm{jam}}), \\
\mathrm{SNR} &= 10\log_{10}(E_{\mathrm{tar}}/E_{\mathrm{n}}).
\end{aligned}
\label{eq:snr_sjr}
\end{equation}
For each sample with specified SNR and SJR levels, the composite jamming profile and noise vector are linearly superposed onto the clean target HRRP template. In this way, HRRP samples contaminated by both noise and composite active jamming can be obtained for subsequent joint recognition.

To support joint target recognition and multi-label jamming classification, each sample is annotated by a unified label vector through binary concatenation:
\begin{equation}
\ell=\bigl[\ell_{\mathrm{tar}},\,\ell_{\mathrm{jam}}\bigr],
\label{eq:label}
\end{equation}
where $\ell_{\mathrm{tar}}\in\{0,1\}^{C_{\mathrm{tar}}}$ denotes the one-hot target label for the corresponding dataset, and $\ell_{\mathrm{jam}}\in\{0,1\}^{4}$ denotes the multi-hot jamming label for the four active jamming types. Here, $C_{\mathrm{tar}}=12$ for the simulation dataset and $C_{\mathrm{tar}}=3$ for the measured dataset. Each element in $\ell_{\mathrm{jam}}$ indicates whether the corresponding jamming mode is present in the current observation.

\begin{figure}[!htbp]
  \centering
  \subfloat[Jamming type distribution]
  {\includegraphics[width=0.48\columnwidth]{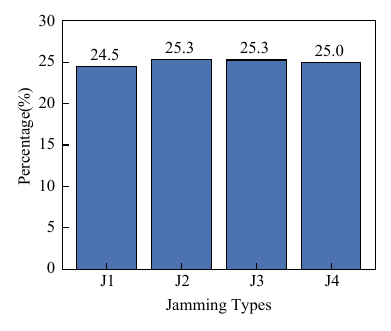}
  \label{fig:dataset_jamming_hist}}
  \hfill
  \subfloat[Label co-occurrence matrix]
  {\includegraphics[width=0.48\columnwidth]{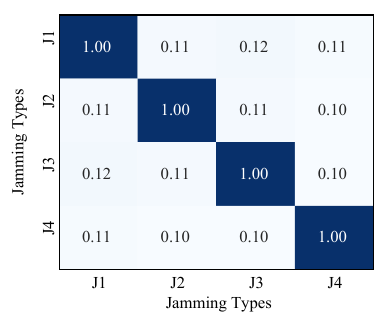}
  \label{fig:dataset_cooccurrence}}
  \caption{Illustrative statistics of the constructed composite jamming dataset built on both the simulation and measured HRRP datasets, including the proportion of each jamming type and the Pearson-correlation-based label co-occurrence matrix with a range of $[-1,1]$. Here, J1--J4 correspond to C\&I, ISRJ, SMSP, and NCJ, respectively. A smaller absolute value indicates weaker correlation between two jamming patterns.}
  \label{fig:dataset_distribution}
\end{figure}

Fig.~\ref{fig:dataset_distribution} shows the statistical distribution of jamming types and the label co-occurrence structure of the constructed composite jamming dataset. The proportions of different jamming types are approximately balanced. Meanwhile, the Pearson-correlation-based co-occurrence matrix shows weak correlations among different jamming labels, indicating that the constructed dataset does not introduce obvious label preference among different jamming combinations. During dataset construction, the SNR and SJR ranges depend on the data source. For the 12-class simulation pipeline, SNR is varied from $-10$ dB to $15$ dB and SJR from $-10$ dB to $0$ dB. For the three-class measured pipeline, SNR is varied from $-10$ dB to $-5$ dB and SJR from $-15$ dB to $-5$ dB. These settings cover typical low-SNR and strong-jamming conditions. The detailed parameter ranges of the four jamming modes are listed in Table~\ref{tab:jamming_params}.

\begin{table}[!htbp]
    \centering
    \caption{Composite jamming parameter ranges.}
    \label{tab:jamming_params}
    \renewcommand{\arraystretch}{1.15}
    \setlength{\tabcolsep}{6pt}
    \setlength{\heavyrulewidth}{1.2pt}
    \setlength{\lightrulewidth}{0.5pt}
    \setlength{\cmidrulewidth}{0.5pt}

    \begin{tabular*}{0.75\columnwidth}{@{\extracolsep{\fill}}l|c@{}}
        \toprule
        \textbf{Parameter} & \textbf{Range} \\
        \midrule
        Jammer bandwidth & 10\,MHz \\
        Velocity & 100--600\,m/s \\
        Range & 1500--5000\,m \\
        Jamming amplitude & 0.8--1.2 \\
        \cmidrule{1-2}
        Sampling duty cycle (ISRJ) & 0.3--0.5 \\
        Number of slices (SMSP) & 2--4 \\
        Frequency step (SMSP) & 1--3\,MHz \\
        Sampling period (C\&I) & 0.5--1.5\,$\mu$s \\
        \bottomrule
    \end{tabular*}
\end{table}

\section{Proposed JointHRRP-Net}
\label{sec:proposed}

\begin{figure*}[!htbp]
  \centering
  \includegraphics[width=0.9\textwidth]{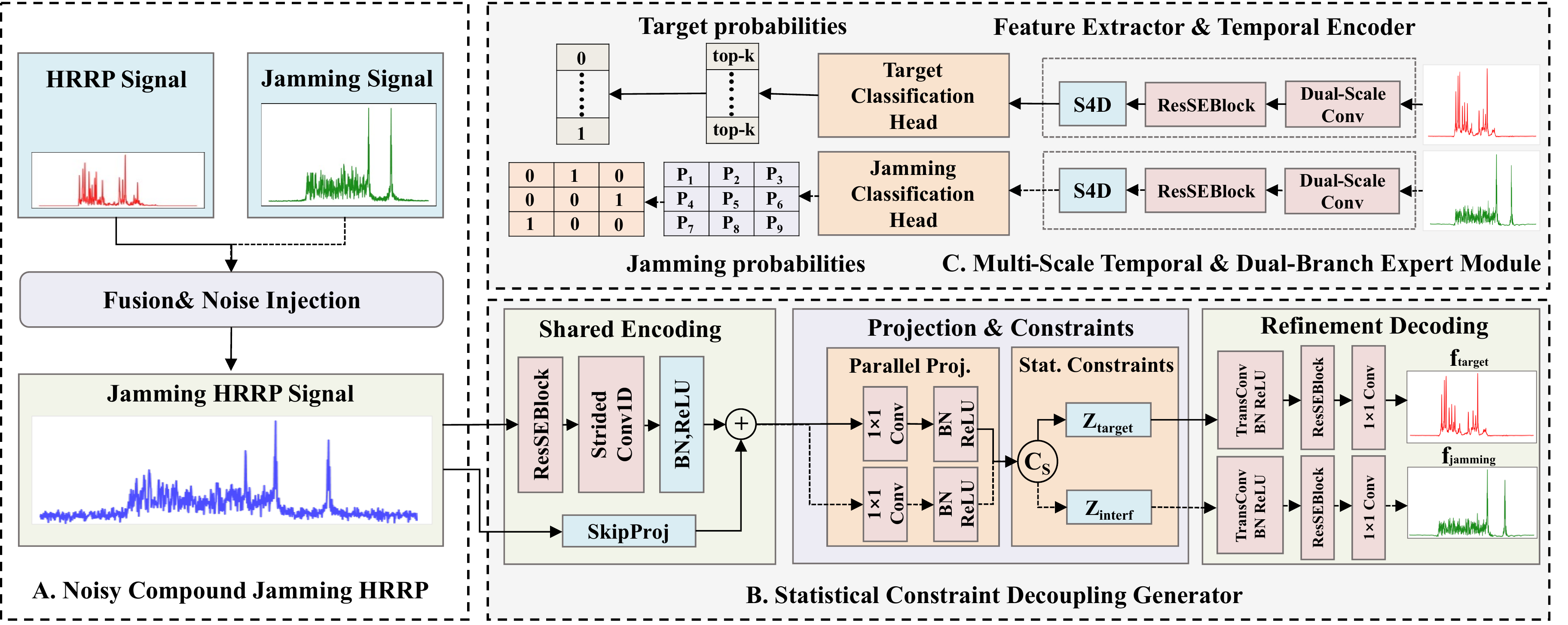}
  \caption{Overall architecture of JointHRRP-Net. (A) Formation of HRRP observations under composite jamming and additive noise. (B) Statistically constrained decoupling module, which generates target-dominant and jamming-dominant latent factors and reconstructs them in the range domain. (C) Multi-scale temporal encoders and dual-expert heads for single-label target classification and multi-label jamming recognition.}
  \label{fig:network_structure}
\end{figure*}

To address target--jamming coupling in HRRP recognition, this paper proposes JointHRRP-Net, a unified framework for HRRP target classification and composite jamming recognition. The overall architecture, illustrated in Fig.~\ref{fig:network_structure}, comprises three main components: a statistically constrained decoupling module, a multi-scale temporal encoding module, and a dual-expert decision module. Specifically, the mixed HRRP profile is first transformed into a shared latent representation, from which target-dominant and jamming-dominant factors are generated through non-shared projections and statistical soft filtering. The two branch features are then processed by multi-scale temporal encoders to capture both local scattering structures and long-range dependencies across HRRP range cells. Finally, task-specific expert heads are adopted to accommodate mutually exclusive single-label target recognition and non-exclusive multi-label jamming recognition within a unified framework.

\subsection{Statistically Constrained Branch Decoupling}
\label{sec:decouple}

\paragraph*{Convolutional stem}
Given an input HRRP profile $\mathbf{x}\in\mathbb{R}^{1\times T}$ with $T$ range cells, a shallow convolutional stem is first used to extract local range-cell responses while preserving the original range resolution:
\begin{equation}
\mathbf{F}_{0}
=
\mathrm{ReLU}
\left(
\mathrm{BN}
\left(
\mathrm{Conv}_{1\times7}(\mathbf{x})
\right)
\right),
\label{eq:stem}
\end{equation}
where $\mathbf{F}_{0}\in\mathbb{R}^{C_{\mathrm{ch}}\times T}$ denotes the stem feature map, $C_{\mathrm{ch}}$ is the channel width, and $\mathrm{BN}$ denotes batch normalization.

\begin{figure}[!htbp]
  \centering
  \includegraphics[width=0.45\textwidth]{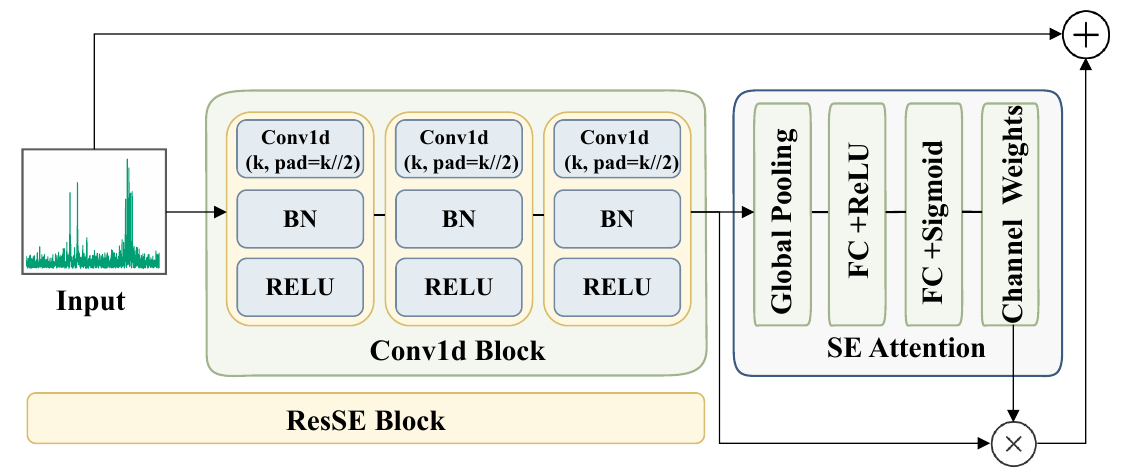}
  \caption{Residual squeeze-and-excitation (ResSE) block. Three stacked 1-D convolutional layers are adopted to enlarge the local receptive field, followed by an SE module \cite{hu2018CVPRSENet} that recalibrates channel-wise responses through global average pooling and fully connected transformations. The reweighted features are then combined with the shortcut connection by element-wise addition. Identity mapping is used when the input and output channels are matched; otherwise, a linear projection is applied.}
  \label{fig:resseblock}
\end{figure}

\paragraph*{Shared compression and branch projection}
The stem feature is compressed into a shared latent representation. The ResSE block shown in Fig.~\ref{fig:resseblock} is used as the basic encoder unit, followed by a strided 1-D convolution. Meanwhile, a parallel $1\times1$ convolution with the same stride serves as a skip projection. The shared latent code is obtained as
\begin{equation}
\mathbf{Z}_{\mathrm{s}}
=
\mathrm{Encoder}(\mathbf{F}_{0})
+
\mathrm{SkipProj}(\mathbf{F}_{0}),
\label{eq:zshared}
\end{equation}
where $\mathbf{Z}_{\mathrm{s}}$ denotes the shared latent code. The skip projection improves gradient propagation and helps preserve low-level range-profile information during compression.

To generate task-oriented latent factors, two non-shared $1\times1$ projections are applied to the shared code:
\begin{equation}
\begin{aligned}
\mathbf{Z}_{\mathrm{tar}}
&=
\mathrm{ReLU}
\left(
\mathrm{BN}
\left(
\mathrm{Conv}^{\mathrm{tar}}_{1\times1}(\mathbf{Z}_{\mathrm{s}})
\right)
\right), \\
\mathbf{Z}_{\mathrm{jam}}
&=
\mathrm{ReLU}
\left(
\mathrm{BN}
\left(
\mathrm{Conv}^{\mathrm{jam}}_{1\times1}(\mathbf{Z}_{\mathrm{s}})
\right)
\right).
\end{aligned}
\label{eq:branchproj}
\end{equation}
where $\mathbf{Z}_{\mathrm{tar}}$ and $\mathbf{Z}_{\mathrm{jam}}$ correspond to target-dominant and jamming-dominant latent factors, respectively. Since both factors are projected from the same mixed representation, a statistical filtering strategy is further introduced to reduce redundant information between the two branches.

\paragraph*{Correlation-guided soft filtering}
Considering the additive formation of HRRP observations under composite jamming, the two projected latent factors are combined as a feature-space mixture proxy:
\begin{equation}
\mathbf{Z}_{\mathrm{mix}}
=
\mathbf{Z}_{\mathrm{tar}}
+
\mathbf{Z}_{\mathrm{jam}}.
\label{eq:zmix}
\end{equation}
  
The proxy provides a latent reference for estimating the correlation between each branch factor and the mixed representation.

For each branch $k\in\{\mathrm{tar},\mathrm{jam}\}$, zero-mean covariance statistics are computed along the range dimension:
\begin{equation}
\begin{aligned}
R_{xy}^{(k)}
&=
\mathrm{Mean}_{t}
\left[
(\mathbf{Z}_{k}-\bar{\mathbf{Z}}_{k})
\odot
(\mathbf{Z}_{\mathrm{mix}}-\bar{\mathbf{Z}}_{\mathrm{mix}})
\right], \\
R_{yy}
&=
\mathrm{Mean}_{t}
\left[
(\mathbf{Z}_{\mathrm{mix}}-\bar{\mathbf{Z}}_{\mathrm{mix}})^2
\right],
\end{aligned}
\label{eq:stats}
\end{equation}
where $\bar{\mathbf{Z}}$ denotes the mean along the range dimension, $\odot$ denotes element-wise multiplication, $R_{xy}^{(k)}$ is the channel-wise cross-covariance between branch $k$ and the mixture proxy, and $R_{yy}$ is the channel-wise variance of the mixture proxy.

The covariance-to-variance ratio is then used to produce a differentiable channel-wise mask:
\begin{equation}
\mathbf{M}_{k}
=
\mathrm{Clamp}
\left(
\frac{R_{xy}^{(k)}}{R_{yy}+\varepsilon},
0,
1
\right),
\label{eq:mask}
\end{equation}
where $\mathbf{M}_{k}$ denotes the soft mask for branch $k$, $\varepsilon$ is a small positive constant, and $\mathrm{Clamp}(\cdot,0,1)$ limits the mask value to $[0,1]$.

The filtered branch factors are obtained by
\begin{equation}
\begin{aligned}
\mathbf{Z}'_{\mathrm{tar}} &= \mathbf{Z}_{\mathrm{tar}}\odot\mathbf{M}_{\mathrm{tar}}, \\
\mathbf{Z}'_{\mathrm{jam}} &= \mathbf{Z}_{\mathrm{jam}}\odot\mathbf{M}_{\mathrm{jam}}.
\end{aligned}
\label{eq:filtered}
\end{equation}

Through this statistical soft filtering, channels with stronger branch-specific consistency are emphasized, while redundant responses shared by the two branches are suppressed. 

\paragraph*{Branch decoding and range-domain reconstruction}
The filtered latent factors are decoded back to the original HRRP resolution. Each branch decoder contains a transposed convolution with stride two and a ResSEBlock:
\begin{equation}
\begin{aligned}
\mathbf{F}_{\mathrm{tar}} &= \mathrm{Decoder}_{\mathrm{tar}}(\mathbf{Z}'_{\mathrm{tar}}), \\
\mathbf{F}_{\mathrm{jam}} &= \mathrm{Decoder}_{\mathrm{jam}}(\mathbf{Z}'_{\mathrm{jam}}),
\end{aligned}
\label{eq:decoder}
\end{equation}
where $\mathbf{F}_{\mathrm{tar}}$ and $\mathbf{F}_{\mathrm{jam}}$ denote the target-dominant and jamming-dominant branch feature maps at full range resolution.

Pointwise $1\times1$ reconstruction heads are further used to map the branch feature maps into range-domain waveforms:
\begin{equation}
\begin{aligned}
\hat{\mathbf{t}} &= \mathrm{Conv}_{1\times1}^{\mathrm{tar}}(\mathbf{F}_{\mathrm{tar}}), \\
\hat{\mathbf{i}} &= \mathrm{Conv}_{1\times1}^{\mathrm{jam}}(\mathbf{F}_{\mathrm{jam}}).
\end{aligned}
\label{eq:recon_heads}
\end{equation}

\subsection{Multi-Scale Temporal Modeling with S4D}
\label{sec:temporal}

After branch decoupling, each stream contains structures with different temporal scales. Localized peaks are usually associated with scattering centers or coherent false targets, whereas broader envelopes are often related to suppression jamming and range-spread jamming. Therefore, a multi-scale temporal encoder is employed to enhance both fine-grained and long-range HRRP representations.

\paragraph*{Micro and macro convolutional stems}
Let $\hat{\mathbf{F}}\in\{\mathbf{F}_{\mathrm{tar}},\mathbf{F}_{\mathrm{jam}}\}$ denote either branch feature map. The temporal encoder contains two parallel convolutional stems. The micro stem uses a $9$-tap convolution with stride four followed by a $17$-tap convolution with stride one, which focuses on local range-cell patterns. In contrast, the macro stem adopts a $127$-tap convolution with stride four to capture wide-range envelope variations. Let $\phi_{k,s}(\cdot)$ denote a 1-D convolution with kernel size $k$ and stride $s$, followed by batch normalization and ReLU activation. The multi-scale fusion is formulated as
\begin{equation}
\begin{aligned}
\mathbf{F}_{\mathrm{micro}}
&=
\phi_{17,1}
\left(
\phi_{9,4}(\hat{\mathbf{F}})
\right), \ 
\mathbf{F}_{\mathrm{macro}}
=
\phi_{127,4}(\hat{\mathbf{F}}), \\
\mathbf{F}_{\mathrm{fused}}
&=
\phi_{1,1}
\left(
\mathrm{ResSEBlock}_{k=7}
\left(
[\mathbf{F}_{\mathrm{micro}},\mathbf{F}_{\mathrm{macro}}]
\right)
\right),
\end{aligned}
\label{eq:multiscale_fusion}
\end{equation}
where $[\cdot,\cdot]$ denotes channel-wise concatenation, $\mathrm{ResSEBlock}_{k=7}(\cdot)$ denotes a seven-tap residual squeeze-and-excitation fusion block, and $\phi_{1,1}(\cdot)$ denotes the final $1\times1$ convolutional projection. 

In this way, the micro stem enhances local scattering details, while the macro stem provides complementary envelope-level information.

\paragraph*{Long-range dependency modeling with S4D}
Although multi-scale convolutions enlarge the receptive field, their dependency modeling ability is still constrained by finite kernel sizes. In HRRP sequences, distant range cells may remain correlated due to extended target scattering, false-target structures, or wideband jamming envelopes. Compared with recurrent models and Transformer self-attention, the diagonal structured state-space model (S4D) provides a more efficient way to model long-range dependencies in one-dimensional sequences. Therefore, an S4D layer is adopted for state-space temporal modeling.

Before S4D modeling, the fused feature map is transposed into sequence-first form and projected to a hidden dimension:
\begin{equation}
\mathbf{U}
=
\mathrm{LN}
\left(
\mathrm{Linear}
\left(
\mathbf{F}_{\mathrm{fused}}^{\top}
\right)
\right),
\label{eq:s4d_input}
\end{equation}
where $\mathbf{U}\in\mathbb{R}^{T'\times H}$ denotes the normalized sequence representation, $T'$ is the downsampled sequence length, $H$ is the hidden width, and $\mathrm{LN}(\cdot)$ denotes layer normalization.

S4D parameterizes a diagonal linear time-invariant state-space model \cite{gu2022iclr_s4,gu2022neurips_s4d,gu2024iclr_mamba}. Its discrete-time form is written as
\begin{equation}
\mathbf{s}_{t+1}
=
\bar{\mathbf{A}}\mathbf{s}_{t}
+
\bar{\mathbf{B}}\mathbf{u}_{t},
\qquad
\mathbf{o}_{t}
=
\mathbf{C}_{\mathrm{out}}\mathbf{s}_{t}
+
\mathbf{D}\mathbf{u}_{t},
\label{eq:s4d_state}
\end{equation}
where $\mathbf{s}_{t}$, $\mathbf{u}_{t}$, and $\mathbf{o}_{t}$ denote the state, input, and output vectors at time step $t$, respectively. $\bar{\mathbf{A}}$ and $\bar{\mathbf{B}}$ are discretized state matrices, while $\mathbf{C}_{\mathrm{out}}$ and $\mathbf{D}$ are the output and feedthrough maps.

The recurrent state-space computation can be equivalently expressed in convolutional form:
\begin{equation}
\mathbf{o}_{t}
=
\mathbf{D}\mathbf{u}_{t}
+
\sum_{\tau=0}^{t}
\mathbf{K}_{t-\tau}\mathbf{u}_{\tau},
\label{eq:s4d_conv}
\end{equation}
where $\mathbf{K}$ is the convolution kernel generated from the diagonal state-space parameters. 

Since this convolution can be evaluated efficiently in the frequency domain, S4D provides an effective trade-off between long-range modeling capability and computational efficiency for one-dimensional HRRP sequences. The S4D output is further processed by GELU activation, dropout, and a $1\times1$ convolutional gated linear unit, producing the sequence feature $\mathbf{F}_{\mathrm{seq}}\in\mathbb{R}^{T'\times H}$.

\paragraph*{Statistical sequence pooling}
To obtain a compact descriptor for each branch, three temporal statistics are aggregated from the S4D output:
\begin{equation}
\mathbf{f}_{\mathrm{pool}}
=
\mathrm{Mean}_{t}(\mathbf{F}_{\mathrm{seq}})
+
\mathrm{Max}_{t}(\mathbf{F}_{\mathrm{seq}})
+
\mathrm{Std}_{t}(\mathbf{F}_{\mathrm{seq}}).
\label{eq:pool}
\end{equation}
Mean pooling summarizes the average response, max pooling retains salient range-cell activations, and standard-deviation pooling describes temporal fluctuation. The same pooling strategy is applied to the target and jamming streams, yielding $\mathbf{f}_{\mathrm{pool}}^{\mathrm{tar}}$ and $\mathbf{f}_{\mathrm{pool}}^{\mathrm{jam}}$, respectively.

\subsection{Dual-Expert Decision Module}
\label{sec:dualhead}

The pooled descriptors are finally fed into two task-specific expert heads. The target expert performs single-label aircraft classification, while the jamming expert performs multi-label prediction over multiple possible jamming modes. This design matches the heterogeneous output spaces of the two tasks without forcing target and jamming decisions into a single homogeneous classifier.

For target recognition, a two-layer multilayer perceptron outputs target logits:
\begin{equation}
\mathbf{o}_{\mathrm{tar}}
=
\mathrm{FC}_{2}
\left(
\mathrm{ReLU}
\left(
\mathrm{FC}_{1}
\left(
\mathbf{f}_{\mathrm{pool}}^{\mathrm{tar}}
\right)
\right)
\right),
\label{eq:target_logits}
\end{equation}
and the target probability vector is obtained by
\begin{equation}
\hat{\mathbf{y}}_{\mathrm{tar}}
=
\mathrm{Softmax}
\left(
\mathbf{o}_{\mathrm{tar}}
\right),
\label{eq:target_head}
\end{equation}
where $\hat{\mathbf{y}}_{\mathrm{tar}}\in\mathbb{R}^{C_{\mathrm{tar}}}$ denotes the predicted target probability vector, and $C_{\mathrm{tar}}$ is the number of target classes.

For jamming recognition, another two-layer expert outputs jamming logits:
\begin{equation}
\mathbf{o}_{\mathrm{jam}}
=
\mathrm{FC}_{2}
\left(
\mathrm{Dropout}
\left(
\mathrm{ReLU}
\left(
\mathrm{FC}_{1}
\left(
\mathbf{f}_{\mathrm{pool}}^{\mathrm{jam}}
\right)
\right)
\right)
\right),
\label{eq:jamming_logits}
\end{equation}
and independent jamming probabilities are calculated by
\begin{equation}
\hat{\mathbf{y}}_{\mathrm{jam}}
=
\mathrm{Sigmoid}
\left(
\mathbf{o}_{\mathrm{jam}}
\right),
\label{eq:jamming_head}
\end{equation}
where $\hat{\mathbf{y}}_{\mathrm{jam}}\in\mathbb{R}^{C_{\mathrm{jam}}}$ denotes the predicted jamming probability vector, and $C_{\mathrm{jam}}$ is the number of jamming types. The sigmoid activation is used because different jamming modes are not mutually exclusive.

\section{Experiments}
\label{sec:experiments}
\subsection{Experimental Setup}
All experiments are conducted on an NVIDIA GeForce RTX 4090D GPU and an Intel Xeon Silver 4214R CPU using Python 3.12 and PyTorch 2.8.0. Training and evaluation use a fixed random seed of 42, and automatic mixed precision is enabled for acceleration. Each composite sample contains one target class, one to four jamming types, and additive white Gaussian noise.
Labels concatenate a $C_{\mathrm{tar}}$-dimensional one-hot target vector and a $C_{\mathrm{jam}}$-dimensional multi-hot jamming vector. The original target prototypes are split into training, validation, and test subsets with a ratio of 75\%:15\%:10\%. 
The training set adopts dynamic sample synthesis in each epoch, while the validation and test sets are constructed with fixed samples to guarantee experimental reproducibility. Unless stated otherwise, tabulated metrics are sample averages on the fixed test set. For fair comparison, all baseline models are retrained on the constructed training set with the same train/validation/test split, HRRP input length, dynamic synthesis strategy, data augmentation protocol, and SNR/SJR sampling ranges as the proposed method. During testing, all models are evaluated on the same fixed composite samples. Parameters and floating-point operations (FLOPs) are reported under the same HRRP input shape for models within each comparison table.

The network is optimized in three stages: decoupling and reconstruction, target classification, and jamming recognition. In each stage, only the corresponding module is updated while the remaining modules are frozen. This setup is chosen to reflect the intended modular design of the framework: during deployment, one can instantiate decoupling only, decoupling with target recognition, or decoupling with jamming recognition, training or substituting only the blocks required for the selected task set. On our benchmark, fully joint end-to-end fine-tuning of all modules achieves closed-set performance comparable to the staged optimization protocol. AdamW and cosine annealing are employed, with the learning rate decaying from $3\times 10^{-3}$ to $10^{-6}$. The training process runs for 100 epochs with an early stopping patience of 20 epochs.

The decoupling stage is supervised directly in the waveform domain. The SI-SNR criterion is introduced to enforce scale-invariant structural consistency between reconstructed and reference components:
\begin{equation}
\label{eq:dec_loss}
\begin{aligned}
\mathcal{L}_{\mathrm{dec}}
&=
-\frac{\lambda_{\mathrm{dec}}}{2}
\left[
\mathrm{SI\text{-}SNR}(\hat{\mathbf{t}},\mathbf{t})
+
\mathrm{SI\text{-}SNR}(\hat{\mathbf{i}},\mathbf{i})
\right], \\
\mathrm{SI\text{-}SNR}(\hat{\mathbf{s}},\mathbf{s})
&=
10\log_{10}
\frac{
\left\|
\displaystyle
\frac{\langle \hat{\mathbf{s}},\mathbf{s}\rangle}
{\|\mathbf{s}\|_2^2+\varepsilon}
\mathbf{s}
\right\|_2^2
}{
\left\|
\hat{\mathbf{s}}
-
\displaystyle
\frac{\langle \hat{\mathbf{s}},\mathbf{s}\rangle}
{\|\mathbf{s}\|_2^2+\varepsilon}
\mathbf{s}
\right\|_2^2
+
\varepsilon
}.
\end{aligned}
\end{equation}
where $\hat{\mathbf{t}}$ and $\hat{\mathbf{i}}$ denote the reconstructed target-dominant and jamming-dominant waveforms, and $\mathbf{t}$ and $\mathbf{i}$ are their corresponding reference components. A larger SI-SNR indicates that the reconstructed waveform is more consistent with its reference after scale projection. The negative sign is therefore used to convert SI-SNR maximization into a minimization objective.

For the target expert, the aircraft categories are mutually exclusive. Therefore, cross-entropy is used for single-label target classification:
\begin{equation}
\mathcal{L}_{\mathrm{tar}}
=
-\log
\frac{
\exp(z_{t,y_t})
}{
\sum_{k=1}^{C_{\mathrm{tar}}}\exp(z_{t,k})
},
\label{eq:target_loss}
\end{equation}
where $z_{t,k}$ is the target logit of class $k$, $C_{\mathrm{tar}}$ is the number of aircraft classes, and $y_t$ is the ground-truth target label.

For the jamming expert, multiple jamming modes may be activated in one observation. Thus, binary cross-entropy with logits is adopted for multi-label jamming recognition:
\begin{equation}
\mathcal{L}_{\mathrm{jam}}
=
-\frac{1}{C_{\mathrm{jam}}}
\sum_{c=1}^{C_{\mathrm{jam}}}
\left[
y_c\log\sigma(z_c)
+
(1-y_c)\log(1-\sigma(z_c))
\right],
\label{eq:jamming_loss}
\end{equation}
where $z_c$ denotes the logit of the $c$th jamming type, $y_c\in\{0,1\}$ is the corresponding ground-truth label, $C_{\mathrm{jam}}$ is the number of jamming types, and $\sigma(\cdot)$ denotes the sigmoid function.

For target recognition, precision, recall, macro F1-score, and confusion matrices are used for evaluation. For jamming recognition, subset accuracy, element-wise accuracy, macro precision, macro recall, and macro F1-score are reported. These metrics provide complementary views for the two tasks, since target recognition follows a single-label decision protocol whereas composite jamming recognition follows a multi-label protocol.

\subsection{Target and Jamming Recognition Performance}

We first evaluated the target recognition performance under composite jamming. Several representative HRRP recognition models are selected as baselines. TACNN incorporates temporal sensing into a CNN backbone \cite{CHEN2022}. CNN-BiRNN combines convolutional feature extraction with bidirectional recurrent modeling \cite{pan2022TGRSHRRPStacked}. The Transformer-based model leverages self-attention to capture long-range dependencies in HRRP sequences \cite{vaswani2017attention,gaotrans}. MSDP-Net performs multi-domain representation learning via multi-scale spatial and spectral feature extraction \cite{msdpnet}. DPFFN further explores staged multi-level feature fusion for radar recognition \cite{zhou2025TAESDualPolHRRP}. All models are evaluated under the same input conditions and jamming settings.

The classification results on the 12-class aircraft electromagnetic simulation dataset are summarized in Table~\ref{tab:target_results}. JointHRRP-Net achieved the best overall performance among all compared methods. Under low-SNR and strong-jamming conditions, the proposed network maintained balanced recognition performance across the 12 aircraft classes, with average precision and recall of approximately 93\%. Compared with the strongest baseline, DPFFN, JointHRRP-Net improved the recognition performance by more than 7 percentage points. Meanwhile, the recognition accuracy of nearly all individual aircraft classes exceeded 90\%, and the lowest class accuracy still reached 89.49\%, demonstrating more stable performance than other comparative models. 

This improvement was mainly attributed to the decoupling-before-recognition design. Instead of directly classifying the mixed HRRP profile, JointHRRP-Net first separated target-dominant and jamming-dominant representations. As a result, the target expert received features with reduced jamming contamination, which improved class separability under composite jamming because the statistical correlation constraints suppressed cross-branch leakage that would otherwise bias the shared representation toward stronger jamming components.

In addition, the performance gain was achieved with moderate computational cost. As shown in Table~\ref{tab:target_results}, JointHRRP-Net had 0.868 M parameters and 0.338 G FLOPs, which remained within a practical deployment range. This indicated that the proposed model provided a favorable balance between recognition accuracy and computational complexity.

\begin{table*}[!htbp]
\centering
\caption{Target recognition on the 12-class simulation dataset. In each accuracy row, the best result is marked in bold and the second-best result is underlined; ties share the same rank. For complexity metrics, smaller values are better.}
\label{tab:target_results}
\small
\renewcommand{\arraystretch}{1.12}
\setlength{\tabcolsep}{4pt}

\begin{tabular*}{\textwidth}{@{\extracolsep{\fill}}llcccccc@{}}
\toprule[1.2pt]
\textbf{Metric group} & \textbf{Target type} & \textbf{TACNN} & \textbf{CNN-BiRNN} & \textbf{Transformer} & \textbf{DPFFN} & \textbf{MSDP-Net} & \textbf{Proposed} \\
\midrule[0.5pt]
\multirow{12}{*}{Accuracy (\%)}
& A1  & 73.52 & 69.14 & 30.56 & \underline{79.05} & 77.91 & \textbf{90.85} \\
& A2  & \underline{99.47} & \textbf{99.85} & 99.24 & 99.39 & 99.17 & \underline{99.47} \\
& A3  & 76.10 & 72.39 & 20.80 & 81.69 & \underline{82.54} & \textbf{93.34} \\
& A4  & 73.98 & 71.79 & 24.51 & \underline{78.97} & 78.52 & \textbf{89.49} \\
& A5  & 79.43 & 77.00 & 49.32 & \underline{83.36} & 83.28 & \textbf{93.12} \\
& A6  & 81.24 & 81.24 & 48.56 & 84.72 & \underline{86.84} & \textbf{94.25} \\
& A7  & 79.95 & 76.85 & 43.12 & 82.60 & \underline{84.42} & \textbf{91.98} \\
& A8  & 75.26 & 72.92 & 42.21 & \underline{84.04} & 81.39 & \textbf{91.91} \\
& A9  & 93.65 & 93.19 & 63.69 & \underline{96.07} & 94.25 & \textbf{97.96}\\
& A10 & 93.04 & 91.00 & 69.67 & 93.12 & \underline{94.10} & \textbf{97.20}\\
& A11 & 83.06 & 77.76 & 33.74 & 83.81 & \underline{83.96} & \textbf{92.89} \\
& A12 & 80.79 & 78.29 & 54.54 & \underline{89.33} & 87.14 & \textbf{95.08} \\
\midrule[0.5pt]
\multirow{3}{*}{Average (\%)}
& Precision & 82.43 & 80.15 & 47.55 & \underline{86.34} & 86.12 & \textbf{93.96}\\
& Recall    & 82.46 & 80.12 & 48.33 & \underline{86.35} & 86.12 & \textbf{93.96} \\
& F1-score  & 82.43 & 80.11 & 47.54 & \underline{86.34} & 86.11 & \textbf{93.96} \\
\midrule[0.5pt]
\multirow{3}{*}{Complexity}
& Parameters (M) & \textbf{0.155} & 1.063 & \underline{0.806} & 1.545 & 0.878 & 0.868 \\
& FLOPs (G)      & \textbf{0.023} & 0.133 & 0.203 & \underline{0.070} & 0.092 & 0.338 \\
& Total test time (s)  & \textbf{1.41}  & \underline{1.84}  & 4.66  & 1.91  & 5.53  & 3.83 \\
\bottomrule[1.2pt]
\end{tabular*}
\end{table*}

\begin{figure}[!htbp]
  \centering
  \subfloat[TACNN]{\includegraphics[width=0.32\columnwidth]{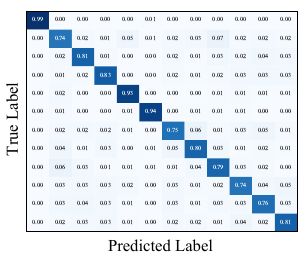}\label{fig:TACNN}}
  \hfill
  \subfloat[CNN-BiRNN]{\includegraphics[width=0.32\columnwidth]{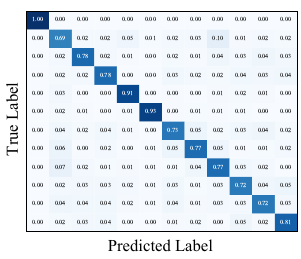}\label{fig:CNNBiRNN}}
  \hfill
  \subfloat[Transformer]{\includegraphics[width=0.32\columnwidth]{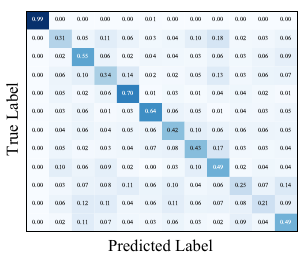}\label{fig:Transformer}}
  \hfill
  \subfloat[DPFFN]{\includegraphics[width=0.32\columnwidth]{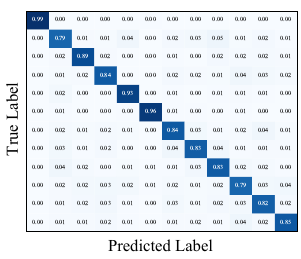}\label{fig:DPFFN}}
  \hfill
  \subfloat[MSDP-Net]{\includegraphics[width=0.32\columnwidth]{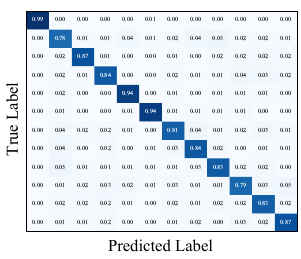}\label{fig:MSDPNet}}
  \hfill
  \subfloat[JointHRRP-Net]{\includegraphics[width=0.32\columnwidth]{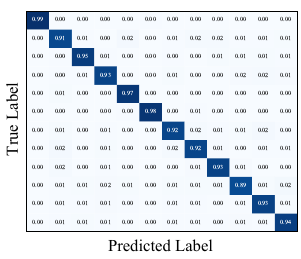}\label{fig:JointHRRPNet}}
  \caption{Confusion matrices for target recognition. Deeper diagonal color indicates higher per-class accuracy. Compared with all baselines, the proposed model exhibits the deepest diagonal entries overall.}
  \label{fig:confusion_matrices}
  \end{figure}

Furthermore, the confusion matrices in Fig.~\ref{fig:confusion_matrices} provided a class-wise comparison on the 12-class simulation dataset. Compared with the baseline models, JointHRRP-Net showed a clearer main diagonal and reduced off-diagonal confusion. This meant that most samples were concentrated in their correct target categories, while misclassification between visually or structurally similar HRRP classes was suppressed. These results were consistent with the quantitative results in Table~\ref{tab:target_results} and indicated that the proposed strategy enhanced target feature separability and improved class-wise discrimination under composite jamming scenarios.

We additionally evaluated target recognition on the three-class measured HRRP dataset, and the results are summarized in Table~\ref{tab:hrrp3data_target_results}. JointHRRP-Net achieved the best overall performance among the compared methods, with class-wise accuracies of about 96\%, 97\%, and 96\% for M1--M3, respectively. In terms of averaged metrics, the proposed model reached about 96\% precision, recall, and F1-score, giving an improvement of roughly 2--3 percentage points over DPFFN. Although TACNN remained competitive on this reduced task, JointHRRP-Net still showed slightly better and more balanced performance across the three measured classes. These results suggested that the proposed decoupling-oriented representation remained effective on measured HRRP data under strong-jamming conditions.

\begin{table*}[!htbp]
\centering
\caption{Target recognition on the three-class measured HRRP dataset. In each performance row, the best result is marked in bold and the second-best result is underlined; ties share the same rank.}
\label{tab:hrrp3data_target_results}
\small
\renewcommand{\arraystretch}{1.12}
\setlength{\tabcolsep}{4pt}

\begin{tabular*}{\textwidth}{@{\extracolsep{\fill}}llcccccc@{}}
\toprule[1.2pt]
\textbf{Metric group} & \textbf{Target type} & \textbf{TACNN} & \textbf{CNN-BiRNN} & \textbf{Transformer} & \textbf{DPFFN} & \textbf{MSDP-Net} & \textbf{Proposed} \\
\midrule[0.5pt]
\multirow{3}{*}{Accuracy (\%)}
& M1 & 95.27 & \underline{95.31} & 82.38 & 93.35 & 93.04 & \textbf{96.35} \\
& M2 & \underline{96.65} & 95.69 & 80.73 & 95.50 & 92.19 & \textbf{97.00} \\
& M3 & 94.19 & 94.46 & 79.96 & 92.77 & \underline{94.85} & \textbf{95.65} \\
\midrule[0.5pt]
\multirow{3}{*}{Average (\%)}
& Precision & \underline{95.38} & 95.17 & 81.07 & 93.88 & 93.56 & \textbf{96.34} \\
& Recall    & \underline{95.37} & 95.15 & 81.03 & 93.87 & 93.36 & \textbf{96.33} \\
& F1-score  & \underline{95.37} & 95.16 & 81.04 & 93.88 & 93.39 & \textbf{96.34} \\
\bottomrule[1.2pt]
\end{tabular*}
\end{table*}

For jamming recognition, extensive experiments were conducted on the 12-class simulation dataset. As shown in Table~\ref{tab:jamming_results}, JointHRRP-Net was compared with four representative multi-label jamming recognition methods: multi-label convolutional neural network (MLCNN) \cite{Zhu2020AMR}, multi-label jamming classification network based on time--frequency maps (MLAMC) \cite{Zhu2020AMR}, multi-feature fusion-based identification network (FRFT-FusionNet) \cite{zhoujam2023}, and hybrid attention recognition network (ShuffleNetV2-EHA) \cite{2024CompoundJammingRecognition}. JointHRRP-Net achieved the best overall performance, with a subset accuracy of 92.97\%, which was approximately 3 percentage points higher than that of the strongest competing model. The recognition accuracy of all four jamming types remained above 97\%, indicating stable discrimination across different jamming patterns. Since subset accuracy requires all active jamming labels in a sample to be predicted correctly, this improvement demonstrated stronger recognition capability under multi-jamming coexistence. The proposed method also obtained a precision of 97.74\%, a recall of 98.68\%, and an F1-score of 98.20\%, further showing that it reduced both missed detections and false alarms.

\begin{table*}[!htbp]
\centering
\caption{Composite jamming recognition performance comparison on the 12-class aircraft electromagnetic simulation dataset. In each performance row, the best result is marked in bold and the second-best result is underlined; ties share the same rank. For complexity metrics, smaller values are better.}
\label{tab:jamming_results}
\small
\renewcommand{\arraystretch}{1.12}
\setlength{\tabcolsep}{4pt}

\begin{tabular*}{\textwidth}{@{\extracolsep{\fill}}llccccc@{}}
\toprule[1.2pt]
\textbf{Metric group} & \textbf{Jamming type} & \textbf{FRFT-FusionNet} & \textbf{ShuffleNetV2-EHA} & \textbf{MLAMC} & \textbf{MLCNN} & \textbf{Proposed} \\
\midrule[0.5pt]
\multirow{4}{*}{Jamming-type accuracy (\%)}
& C\&I & 88.60 & 97.16 & 95.80 & \underline{97.18} & \textbf{97.76}\\
& ISRJ & 88.02 & 95.78 & 93.27 & \underline{96.33} & \textbf{97.16}\\
& SMSP & 84.71 & \underline{96.52} & 93.25 & 95.86 & \textbf{97.42} \\
& NCJ  & 91.52 & \underline{97.93} & 97.28 & 97.62 & \textbf{98.62} \\
\midrule[0.5pt]
\multirow{5}{*}{Average (\%)}
& Subset accuracy  & 63.10 & \underline{89.67} & 83.02 & 89.40 & \textbf{92.97} \\
& Element-wise accuracy & 88.21 & \underline{96.85} & 94.90 & 96.75 & \textbf{97.74} \\
& Precision        & 87.82 & \underline{96.93} & 95.38 & 96.29 & \textbf{97.74} \\
& Recall           & 94.56 & 98.07 & 96.65 & \underline{98.63} & \textbf{98.68} \\
& F1-score         & 90.98 & \underline{97.50} & 95.96 & 97.44 & \textbf{98.20} \\
\midrule[0.5pt]
\multirow{3}{*}{Complexity}
& Parameters (M) & \underline{0.657} & \textbf{0.648} & 33.781 & 2.993 & 1.567 \\
& FLOPs (G)      & \underline{0.166} & \textbf{0.037} & 0.652 & 0.762 & 0.513 \\
& Total test time (s) &\underline{5.33} & \textbf{1.99} & 4.05 & 3.06 & 3.75 \\
\bottomrule[1.2pt]
\end{tabular*}
\end{table*}

\subsection{Robustness and Open-Set Evaluation}

\begin{figure*}[!htbp]
  \centering
  \subfloat[Target accuracy versus SNR]{\includegraphics[width=0.5\columnwidth]{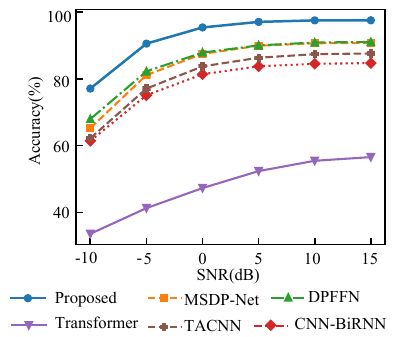}\label{fig:targetsnr}}
  \hfill
  \subfloat[Jamming subset accuracy versus SNR]{\includegraphics[width=0.5\columnwidth]{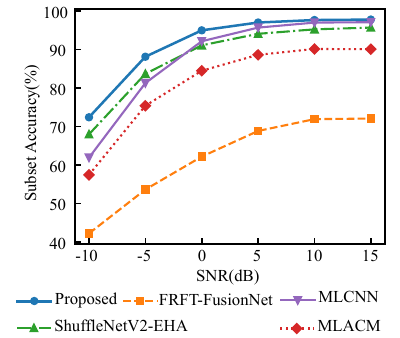}\label{fig:subset_accuracy_vs_snr}}
  \hfill
  \subfloat[Target accuracy versus SJR]{\includegraphics[width=0.5\columnwidth]{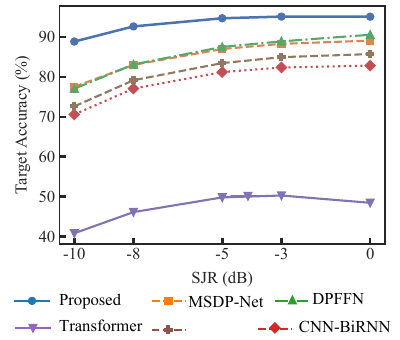}\label{fig:target_accuracy_vs_sjr}}
  \subfloat[Jamming subset accuracy versus SJR]{\includegraphics[width=0.5\columnwidth]{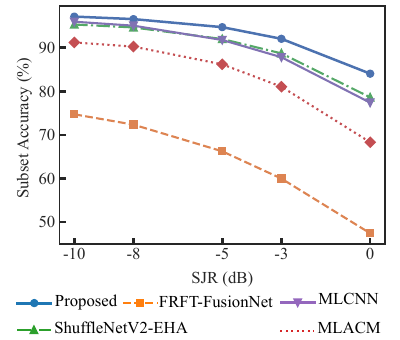}\label{fig:interference_subset_accuracy_vs_sjr}}

  \caption{Robustness evaluation of the proposed model under varying SNR and SJR conditions on the 12-class simulation dataset.}
  \label{fig:snr_performance}
\end{figure*}

To evaluate the robustness of the proposed method under varying noise and jamming intensities, extensive experiments were conducted on the 12-class simulation dataset. Target recognition accuracy and jamming subset accuracy were tested under different SNR and SJR conditions, as shown in Fig.~\ref{fig:snr_performance}. Under low-SNR conditions, JointHRRP-Net maintained approximately 80\% target recognition accuracy at $-10$ dB and exceeded 90\% when the SNR was higher than $-5$ dB. Meanwhile, its jamming subset accuracy remained consistently higher than that of the compared methods and reached approximately 95\% when the SNR exceeded 0 dB. These results indicated that the proposed model preserved both target discrimination and jamming recognition capability under noisy observations.

Under varying SJR conditions, all methods suffered from performance degradation as jamming became stronger. However, JointHRRP-Net showed the smallest accuracy drop and maintained target recognition accuracy close to 90\% under $-10$ dB SJR. For composite jamming recognition, the proposed model also achieved the most stable subset accuracy across the whole SJR range. This demonstrated that the decoupling-oriented design mitigated jamming effects on target classification while providing robust jamming-specific representations.

\begin{figure}[!htbp]
  \centering
  \subfloat[Target recognition accuracy]{\includegraphics[width=0.5\columnwidth]{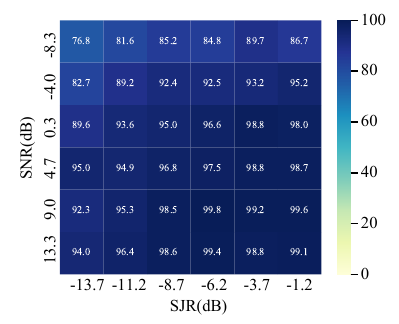}\label{fig:heat1}}
  \hfill
  \subfloat[Jamming recognition subset accuracy]{\includegraphics[width=0.5\columnwidth]{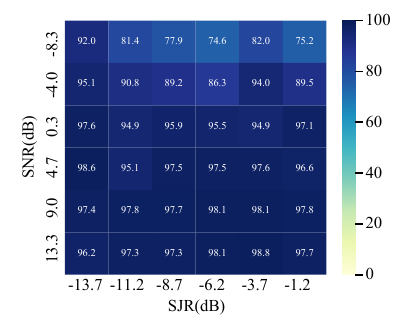}\label{fig:heat2}}
  \caption{Recognition accuracy heatmaps under varying SNR and SJR conditions. The color intensity corresponds to recognition performance, with deeper colors indicating higher accuracy. These heatmaps illustrate the model's performance across different noise and jamming intensity combinations.}
 \label{fig:snr_sjr}
  \end{figure}

To further examine the joint influence of noise and jamming intensity on the 12-class simulation dataset, Fig.~\ref{fig:snr_sjr} presents the heatmaps of target recognition accuracy and jamming recognition accuracy under different SNR--SJR combinations. The proposed model maintained reliable performance over a wide range of operating conditions, including the low-SNR and low-SJR regions. In particular, at the most challenging tested boundary where both SNR and SJR reached $-10$ dB, JointHRRP-Net still achieved approximately 76\% target recognition accuracy and 92\% jamming recognition accuracy. This result indicated that the proposed model preserved discriminative target and jamming representations under simultaneous severe noise and strong jamming, thereby providing useful recognition support for radar perception in complex electromagnetic environments.

For open-set evaluation, OpenMax \cite{bendale2016towards} was applied to the closed-set target expert features. This setting was used to examine whether the target representations learned under composite jamming could also support unknown-target rejection, which is consistent with recent HRRP-oriented open-set frameworks and prototype regularization methods \cite{li2025TAESOSFSM,chen2025TAESECAPL,zhang2025TAESHRRPSeqNet}. The twelve aircraft classes were divided into nine known classes and three unknown classes in a fixed order. For each known class, validation samples that were correctly classified were used to estimate the class centroid and fit the corresponding Weibull tail distribution.

\begin{figure}[!htbp]
\centering
\includegraphics[width=0.35\textwidth]{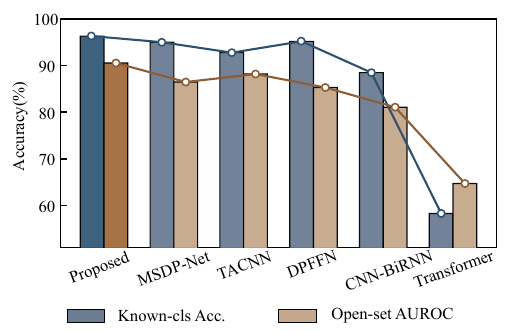}
\caption{Comparison of closed-set accuracy and open-set AUROC across all models after OpenMax post-processing. Each model is represented by a pair of bars, connected by a line to indicate the trend. The proposed model achieves the best trade-off between the two metrics.}
\label{fig:open_set}
\end{figure}

During testing, OpenMax performed top-3 score recalibration with a rejection threshold of 0.5. A class-conditional distance metric combining Euclidean distance and cosine distance was adopted for tail fitting and unknown rejection. Under the same composite jamming conditions as the closed-set experiments, JointHRRP-Net achieved 96.29\% known-class accuracy, 99.75\% known-class acceptance rate, and 90.53\% open-set AUROC. As shown in Fig.~\ref{fig:open_set}, the proposed model outperformed the competing architectures in open-set evaluation. These results indicated that the learned target embeddings maintained good separability between known and unknown targets, even when the HRRP observations were contaminated by composite jamming.

\subsection{Decoupling Performance Verification}

\begin{figure*}[!htbp]
  \centering
  \setlength{\tabcolsep}{1.5pt}
  \renewcommand{\arraystretch}{0.98}
 \begin{tabular}{@{}cccc@{}}
  \includegraphics[width=0.24\textwidth]{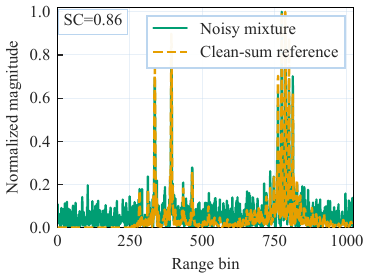} &
  \includegraphics[width=0.24\textwidth]{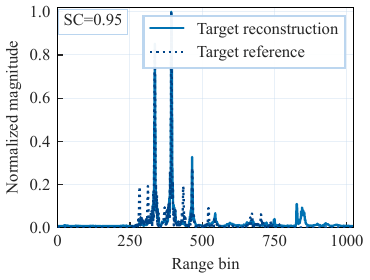} &
  \includegraphics[width=0.24\textwidth]{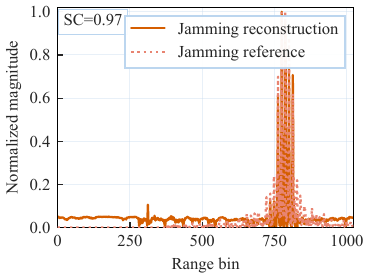} &
  \includegraphics[width=0.24\textwidth]{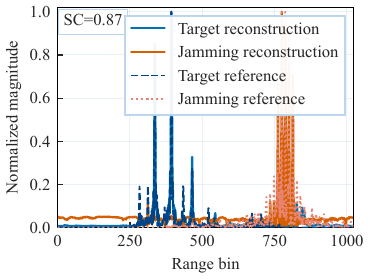} \\
  \includegraphics[width=0.24\textwidth]{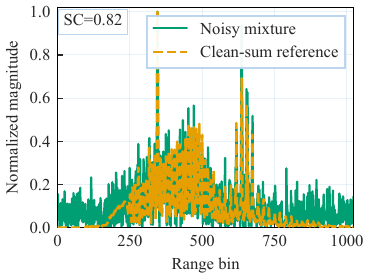} &
  \includegraphics[width=0.24\textwidth]{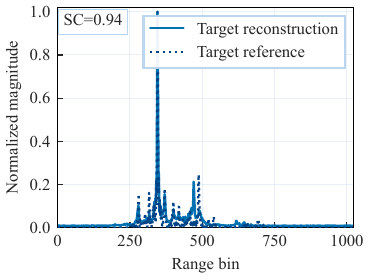} &
  \includegraphics[width=0.24\textwidth]{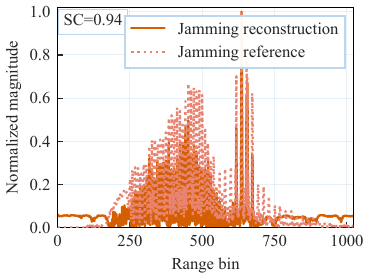} &
  \includegraphics[width=0.24\textwidth]{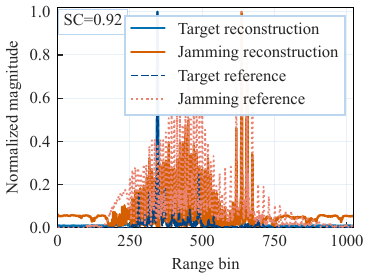} \\
  \includegraphics[width=0.24\textwidth]{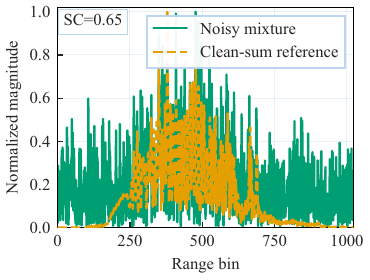} &
  \includegraphics[width=0.24\textwidth]{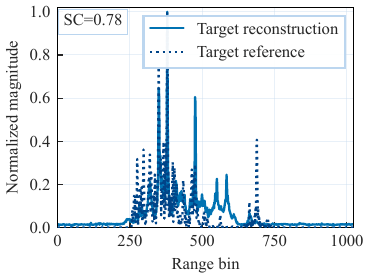} &
  \includegraphics[width=0.24\textwidth]{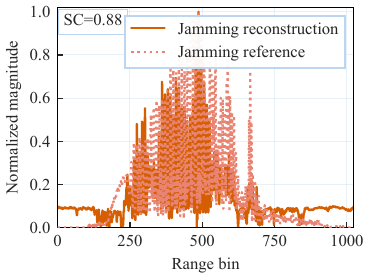} &
  \includegraphics[width=0.24\textwidth]{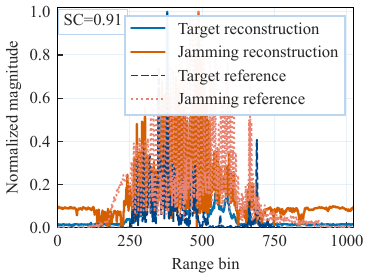} \\
  \includegraphics[width=0.24\textwidth]{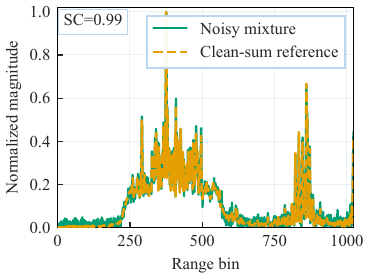} &
  \includegraphics[width=0.24\textwidth]{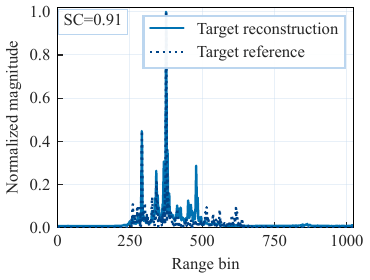} &
  \includegraphics[width=0.24\textwidth]{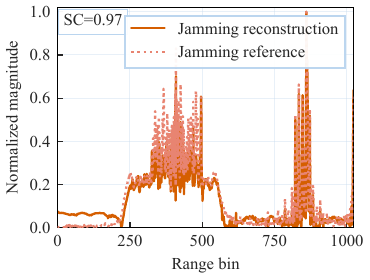} &
  \includegraphics[width=0.24\textwidth]{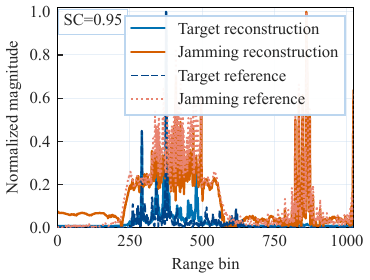} \\
  \footnotesize (a) & \footnotesize (b) & \footnotesize (c) & \footnotesize (d)
  \end{tabular}
  \caption{Decoupling visualization atlas under different jamming complexities. Rows correspond to the number of active jamming types $|J|\in\{1,2,3,4\}$. Columns show: (a) noisy mixture versus noise-free target--jamming sum, (b) reconstructed target component versus clean target reference, (c) reconstructed jamming component versus clean jamming reference, and (d) recomposed signal $\hat{t}+\hat{j}$ versus clean composite reference $t+j$. Curves denote normalized magnitude versus range bin. Each panel shows one target-correct example, and the inset reports the structural correlation coefficient (SC), measured by Pearson correlation.}
  \label{fig:decoupling}
\end{figure*}

To verify the separation capability of the decoupling module under different jamming-combination complexities, Fig.~\ref{fig:decoupling} presents a decoupling visualization atlas with the number of active jamming types $|J|$ varying from 1 to 4. The figure compares the noisy mixed inputs, branch-wise reconstructions, and recomposed signals with their corresponding references.

The reconstructed target and jamming components preserved the main waveform structures of their clean references across different values of $|J|$. Quantitatively, the average structural correlation coefficient (SC) reached approximately 0.91 for the target component and approximately 0.96 for the jamming component. In addition, the recomposed signal $\hat{t}+\hat{j}$ remained consistent with the clean composite reference $t+j$, indicating that the decoupling module separated target-dominant and jamming-dominant components while retaining the principal structural information of the mixed HRRP observation.

\subsection{Ablation Study}

To analyze the contribution of each key component, ablation experiments were conducted on the 12-class simulation dataset. Several controlled ablation variants were evaluated. The configurations included the proposed full JointHRRP-Net, a variant without statistical correlation filtering, a variant replacing the decoupling module with a simplified autoencoder (AE), a variant without S4D, and a variant without the decoupling module. The evaluation metrics comprised target overall accuracy (OA), jamming subset accuracy, target reconstruction cosine similarity, and branch leakage.

To quantify the overlap between the reconstructed target and jamming components, the leakage metric is defined as
\begin{equation}
\mathrm{Leakage}
=
\frac{
\mathrm{Mean}\left(|\hat{\mathbf{t}}\odot\hat{\mathbf{i}}|\right)
}{
\sqrt{
\mathrm{Mean}(\hat{\mathbf{t}}^2)
\mathrm{Mean}(\hat{\mathbf{i}}^2)
}
},
\label{eq:leakage}
\end{equation}
where $\hat{\mathbf{t}}$ and $\hat{\mathbf{i}}$ denote the reconstructed target and jamming components, respectively, $\odot$ denotes element-wise multiplication, and $\mathrm{Mean}(\cdot)$ computes the arithmetic mean over all entries. A smaller leakage value indicates weaker cross-branch overlap and better component separation.

The ablation results are summarized in Table~\ref{tab:ablation}. The proposed model achieved the best performance across all metrics, with a target OA of 93.96\%, a jamming subset accuracy of 92.97\%, a target cosine similarity of 91.19\%, and the lowest leakage value of 0.473. This result indicated that the complete design provided the most effective balance between recognition accuracy, reconstruction consistency, and branch separation.

Removing the statistical correlation filtering reduced the target OA from 93.96\% to 91.99\% and increased the leakage from 0.473 to 0.491. This degradation suggested that the covariance-based soft filtering helped suppress redundant information between the target and jamming branches, because the filtering step explicitly attenuated cross-branch correlation in the latent space before reconstruction. When the decoupling module was replaced by a simplified AE, the target cosine similarity dropped sharply to 44.26\%, while the leakage increased to 0.765. This indicated that a simple reconstruction structure was insufficient to preserve target-related waveform characteristics and separate mixed components effectively.
The variant without S4D showed the largest decrease in target OA, falling to 83.42\%. This result demonstrated the importance of long-range temporal modeling for HRRP target recognition under strong jamming. In contrast, removing the decoupling module entirely led to a target OA of 89.40\% and a jamming subset accuracy of 91.44\%. Since this variant did not produce separated target and jamming components, the target cosine similarity and leakage metrics were not applicable. Overall, these ablation results showed that statistical soft filtering, branch decoupling, and S4D-based temporal modeling jointly contributed to robust recognition under composite jamming.

\begin{table}[!htbp]
\caption{Ablation study of key components. The best value in each column is marked in bold, and the second-best value is underlined. For leakage, lower values indicate better separation.}
\label{tab:ablation}
\centering
\small
\setlength{\tabcolsep}{3.5pt}%
\begin{tabular*}{\columnwidth}{@{\extracolsep{\fill}}lcccc@{}}
\toprule
Model variant & \makecell{Target\\OA (\%)} & \makecell{Subset\\acc. (\%)} & \makecell{Target\\cosine (\%)} & Leakage \\
\midrule
Proposed              & \textbf{93.96} & \textbf{92.97} & \textbf{91.19} & \textbf{0.473} \\
Simple AE             & 91.38 & 91.89 & 44.26 & 0.765 \\
Without stat. corr.   & \underline{91.99} & \underline{91.94} & \underline{89.98} & \underline{0.491} \\
Without S4D           & 83.42 & 91.67 & 88.64 & 0.542 \\
Without decoupling    & 89.40 & 91.44 & --    & -- \\
\bottomrule
\end{tabular*}
\end{table}

\begin{figure}[!htbp]
  \centering
  \includegraphics[width=0.57\columnwidth]{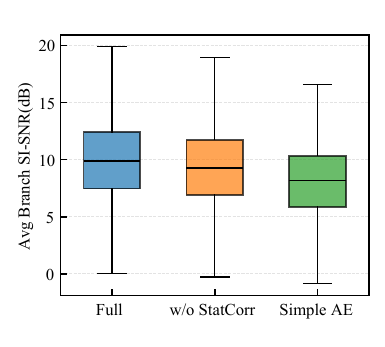}
  \caption{Distribution of average SI-SNR across ablation configurations. The proposed model achieves the highest median, indicating better reconstruction consistency and component separation.}
  \label{fig:SI-SNR}
\end{figure}

Furthermore, Fig.~\ref{fig:SI-SNR} compared the SI-SNR distributions of the reconstructed components under different ablation configurations. The proposed model achieved the highest median and a more concentrated distribution, indicating more stable waveform reconstruction for the decoupled target and jamming branches. Removing the statistical correlation filtering led to an overall decrease in SI-SNR, which further supported the role of statistical soft filtering in improving branch separation. The simplified AE showed a more significant decline, suggesting that the complete decoupling architecture was necessary for reliable component reconstruction. These observations were consistent with results in Table~\ref{tab:ablation}, verifying the effectiveness of the proposed decoupling design.

\section{Conclusion}

In this paper, JointHRRP-Net was proposed for unified HRRP target recognition and composite jamming recognition under complex electromagnetic interference. The proposed framework first performed target--jamming decoupling through a statistically constrained decoupling module, in which parallel projections and correlation-guided soft filtering were employed to separate target-dominant and jamming-dominant representations. Then, multi-scale convolutional stems and an S4D-based temporal encoder were integrated to capture local scattering structures and long-range range-cell dependencies. Finally, dual-expert heads were designed to accommodate the heterogeneous outputs of single-label target classification and multi-label jamming recognition.
Extensive experiments demonstrated the effectiveness of the proposed framework. Compared with representative HRRP recognition and jamming recognition methods, JointHRRP-Net achieved better target recognition accuracy and more stable composite jamming recognition under low-SNR and low-SJR conditions. The robustness evaluation further showed that the model maintained reliable performance when noise and jamming intensity varied simultaneously. In addition, open-set evaluation indicated that the learned target representations remained discriminative for unknown-target rejection, while visualization and ablation studies verified the effectiveness of the decoupling module and the S4D-based temporal modeling strategy.
A current limitation is that the present framework focuses on single-target observations under composite jamming. Future work will consider recognition under multi-target and multi-jamming conditions, together with transfer learning and domain adaptation strategies to reduce annotation dependence and enhance generalization to more practical radar operating conditions.

\bibliographystyle{IEEEtran}
\bibliography{refieee}

\end{document}